\documentclass[preprint]{aastex}

\usepackage{amssymb}
\usepackage{amsmath}
\usepackage{graphicx}
\usepackage{epsfig}
\usepackage{enumitem}
\usepackage{verbatim}
\usepackage[position=t,singlelinecheck=false,caption=false]{subfig}

\begin{document}

\title{THE IMPORTANCE OF JET BENDING IN GAMMA-RAY AGN - REVISITED}

\author{P. J. Graham}
\affil{School of Physics, University of New South Wales, Sydney, NSW 2052, Australia}
\affil{International Centre for Radio Astronomy Research, Curtin University, Bentley, WA, Australia}
\and
\author{S. J. Tingay}
\affil{International Centre for Radio Astronomy Research, Curtin University, Bentley, WA, Australia}

\begin{abstract}

\noindent We investigate the hypothesis that $\gamma$-ray-quiet AGN have a larger tendency for jet bending than $\gamma$-ray-loud AGN, revisiting the analysis of \citet{Tingay1998}. We perform a statistical analysis using a large sample of 351 radio-loud AGN along with $\gamma$-ray identifications from the Fermi Large Area Telescope (LAT). Our results show no statistically significant differences in jet-bending properties between $\gamma$-ray-loud and $\gamma$-ray-quiet populations, indicating that jet bending is not a significant factor for $\gamma$-ray detection in AGN.

\end{abstract}


\section{Introduction}

The Energetic Gamma-Ray Experiment Telescope (EGRET) detected 66 active galactic nuclei (AGN) with $\gamma$-ray emission at energies greater than 100 MeV \citep{Mattox1997a,Hartman1999}. Almost all of the AGN identified were blazars, triggering immense interest and prompting a series of studies focused on the multi-wavelength properties of these AGN. EGRET's successor, the Large Area Telescope (LAT, onboard the \textit{Fermi Gamma-Ray Space Telescope}), has identified over 1000 $\gamma$-ray sources with energies greater than 100 MeV \citep{Ackermann2011b}, revealing the $\gamma$-ray sky in detail. The LAT instrument presents the opportunity to revisit questions first examined in the EGRET era with far superior data.

EGRET's discovery opened a new direction in investigating the nature and physical processes of blazars. Blazars are radio sources characterized by a compact core and relativistic jets that are aligned at a small angle to our observational line-of-sight, i.e. flat-spectrum radio quasars and BL Lac objects \citep{Fichtel1994,Impey1996,Mattox1997b}. It is supposed that $\gamma$-ray emission from AGN is produced from inverse-Compton (IC) processes occuring within the jets and becomes apparently enhanced via relativistic beaming \citep{Mattox1997b}. Although there is good support for this scenario, several questions remain, particularly with respect to the origin of $\gamma$-ray production within the jets \citep{Jorstad2001,Lahteenmaki2003,Sikora2009}, as well as how IC processes act to produce $\gamma$-ray emission \citep{Sikora2009}.

Relativistic beaming (or Doppler boosting) is evidenced by other associated properties that characterize blazars: high $\gamma$-ray luminosity (which comprises a large portion of the total radiative output) \citep{Mattox1997b}; apparent superluminal motion \citep{Jorstad2001}; and rapid variability from radio to $\gamma$-ray wavelengths \citep{Aller1996,Lahteenmaki2003,Kovalev2009}. From the non-thermal nature of this emission, it is understood that it is produced from within the AGN jets and is relativistically beamed along the line-of-sight. This could explain why $\gamma$-ray-loud AGN are also radio-loud but not all radio-loud AGN are $\gamma$-ray-loud. If $\gamma$-ray emitting regions move at faster speeds than radio emitting regions, $\gamma$-ray emission would be Doppler boosted within a narrower cone \citep{Salamon1994}. Outside the $\gamma$-ray beaming cone and within the radio beaming cone, $\gamma$-ray radiation would be Doppler dimmed, giving the appearance of a $\gamma$-ray-quiet yet radio-loud AGN.

A second possibility suggests that the appearance of $\gamma$-ray emission may be related to AGN jet-bending \citep{vonMontigny1995}. If a bend in the jet occurs downstream of the $\gamma$-emitting region and upstream of regions of extended radio emission, the emissions would be Doppler-boosted in alignment to their respective sections of jet. As such, it may be possible that $\gamma$-ray emission is beamed away from us whilst radio emission is beamed towards us due to a better alignment of the radio-emitting jet section with our line-of-sight.

Several studies have compared $\gamma$-ray-loud and $\gamma$-ray-quiet AGN populations with respect to characteristics of the AGN core and jet components. Statistically significant differences between these populations are found for core brightness temperature, jet opening angles and core polarization \citep{Mattox1997b,Tingay2002,Taylor2007,Pushkarev2009,Linford2011}. The distinction between populations is also established for radio flux density variability between $\gamma$-ray-loud and $\gamma$-ray-quiet AGN \citep{Aller1996,Lahteenmaki1999,Tingay2003,Kovalev2009,Richards2011}, suggesting that radio and $\gamma$-ray emission are produced within the same region within AGN jets \citep{Lahteenmaki1999}. \citet{Hughes2011} propose that radio-to-$\gamma$-ray variability is caused by oblique shocks in the AGN jets. The role of oblique shocks has also been explored in jets on parsec \citep{Tingay1996} and kiloparsec scales \citep{Balsara1992,Lebedev2004}. \citet{Ackermann2011a} finds a correlation between $\gamma$-ray and radio flux in AGN after consideration of biases unaccounted for in previous studies \citep{Mucke1997}. Studies have also correlated superluminal ejections and $\gamma$-ray flares in AGN jets, suggesting that $\gamma$-ray emission is produced within the parsec-scale region of the jets \citep{Jorstad2001,Lahteenmaki2003,Kovalev2009,Pushkarev2010}. Furthermore, studies have related $\gamma$-ray emission with jet kinematics \citep{Kellermann2004,Lister2009b,Savolainen2010}, and there is evidence to suggest that FSRQs and BL Lac objects may have intrinsically different mechanisms for $\gamma$-ray emission \citep{Ackermann2011b,LeonTavares2011,Lister2011,Nieppola2011,Linford2012}. 

In examining the jet bend scenario of \citet{vonMontigny1995}, \citet[hereafter Paper I]{Tingay1998} used EGRET identifications to compare $\gamma$-ray-loud and $\gamma$-ray-quiet AGN with respect to their jet-bending characteristics. Their results showed a tendency in $\gamma$-ray-quiet sources to both have more jet bends and more pronounced jet-bend angles at parsec scales. \citet{Bower1997} did not find evidence to support this scenario, but their study was limited in only examining parsec-to-kiloparsec jet misalignment angles. The jet bend proposition was revisited in \citet{Taylor2007} and \citet{Linford2011} with inconclusive results (limited to 4 EGRET candidates and 30 LAT sources respectively). Similarly, \citet{Linford2012} found that no significant difference in jet-bend angles between $\gamma$-ray-loud and $\gamma$-ray-quiet populations (limited to 19 LAT sources).

Utilizing LAT's superior $\gamma$-ray detection capabilities and a larger sample size, we re-examine the suggestion that jet-bending may be a significant factor for the detection of $\gamma$-ray emission in AGN. We follow the analysis of Paper I, testing the statistical significance of jet-bend angles, as well as the number of bends between $\gamma$-ray-loud and $\gamma$-ray-quiet AGN. We use the `Clean Sample' of the second LAT AGN catalog \citep[2LAC,][]{Ackermann2011b} and VLBI images accessible in the published literature or from the Radio Fundamental Catalog (RFC, version `rfc\textunderscore2013d'). We examine the statistics of two samples, a large but inhomogeneous sample and a smaller but homogeneous subsample based on \citet[hereafter MOJAVE]{Lister2009a}.

\section{Sample Definitions}

In order to investigate the significance of jet-bending between $\gamma$-ray-loud and $\gamma$-ray-quiet AGN populations, we surveyed the literature for VLBI images and measured source properties.

\subsection{The RFC-Based Sample}

The most recent version of the RFC contains 8310 compact radio sources studied in 5719 VLBI observing sessions, making it the most complete catalog of radio sources with available VLBI data. We have compared the RFC with the CRATES catalog \citet{Healey2007}, a nearly uniform survey of of flat-spectrum radio sources ($\alpha > -0.5$) containing flux densities and spectral indices for over 11,000 objects. By cross-referencing data from the CRATES catalog to sources in the RFC, we obtain a large pool of sources to apply selection criteria and draw upon in order to define a `RFC-Based Sample' for the purpose of our investigation.

There are a number of considerations in formulating our selection criteria for the RFC-Based Sample. A true cross-identification between radio/X-ray and $\gamma$-ray counterparts can only be determined when there is a correlation in variability in both bands. In practice, cross-identifications were made only for 28 sources in the 2-Year Fermi-LAT Sources Catalog \citep[2FGL,][]{Nolan2012}. As such, most sources identified as $\gamma$-ray-loud in 2LAC are high-confidence statistical associations determined by three association methods using the detected positions and uncertainties of radio/X-ray and $\gamma$-ray emissions within a source's 95\% error ellipse \citep{Ackermann2011b}, along with the physical properties of likely candidates.

The determination of a high-confidence association in 2LAC is also subject to a number of considerations \citep{Ackermann2011b,Nolan2012}. Candidate gamma-ray sources are `detected' in 2FGL if a region of concentrated flux against its background obtains a test statistic TS $>$ 25 (corresponding to over 4 $\sigma$ significance). There is low availability of spectroscopic information for radio sources in southern declinations. There are in some case multiple associations within a gamma-ray source's 95\% error ellipse. A candidate's gamma-ray spectral intensity may not lie within LAT's energy-dependent sensitivity curve.  Candidates with weak radio flux would also have reduced likelihoods of association as they become indistinguishable from other weak radio sources within the LAT error ellipse. These factors would cause bias to higher unassociated AGN despite possibly having detectable $\gamma$-ray flux. To address these issues, we employ the `Clean Sample' subset of 2LAC where sources are excluded if they presented difficulties in the analysis of making an association. We also apply a low threshold to flux densities to account for the possibility of false non-associations below the cutoff. 

There is also possible bias introduced by $\gamma$-ray attenuation due to the Extragalactic Background Light (EBL), resulting in a lower rate of association at higher redshifts. Predictions from EBL models show that $\gamma$-rays with energies below $\sim$ 10 GeV from redshifts up to z $\sim$ 3 do not undergo significant attenuation \citep{Stecker2006,Franceschini2008,Abdo2010}, and \citet{Abdo2010} find no redshift dependence in the flux ratio of $\gamma$-ray photons F($>$ 10 GeV)/F($>$ 1 GeV) across blazar subclasses from z = 0 to above z = 2. Applying an upper limit to redshift can eliminate the possibility of this bias in our sample.

Furthermore, errors in modelling the diffuse $\gamma$-ray background near the Galactic ridge or in nearby interstellar cloud regions may introduce false associations between $\gamma$-ray and radio emissions belonging to two distinct sources, identifying `$\gamma$-ray-loud' AGNs that are actually $\gamma$-ray-quiet. However, 2LAC selection criteria include only sources with high Galactic latitudes ($|b| > 10^{\circ}$), eliminating the possibility of false or non-associations occuring near the Galactic ridge.


In review of these factors in $\gamma$-ray detection and source association, our source selection criteria are therefore as follows:

\begin{enumerate}
	\item identified in RFC;
	\item Galactic latitude $|b| > 10^{\circ}$;
	\item total 4.85 GHz flux density $S_{4.85} > 0.8$ Jy;
	\item redshift z $<$ 2;
	\item spectral index $\alpha_{4.85/low} > -0.5$;
	\item included in Clean Sample if included in 2LAC;
	\item associated with VLBI images in literature.
\end{enumerate}

\noindent These criteria define our RFC-Based Sample, consisting of 351 AGN in total. The sample contains 151 $\gamma$-ray-loud AGN from the Clean Sample, a subset of sources from the second LAT catalog that possess high-confidence associations between radio and $\gamma$-ray counterparts \citep{Ackermann2011b} (see Table \ref{tab:gl_list}). A $\gamma$-ray-quiet subsample was defined by the 200 AGN not detected by LAT (see Table \ref{tab:gq_list}).

\subsection{The MOJAVE Subample}

We consider a subsample of the RFC-Based Sample defined by sources existing in MOJAVE to investigate concerns of sample inhomogeneity \citep{Lister2009a}. The `MOJAVE Subsample' is an unbiased subset since the sources included conform to our selection criteria for the RFC-Based Sample. The sample contains 65 $\gamma$-ray-loud AGN and 29 $\gamma$-ray-quiet AGN, giving a total sample of 94 AGN. The sources were observed using the VLBA and processed uniformly. The selection criteria used in MOJAVE are also relatively homogeneous, only departing slightly from uniformity in the criterion for radio flux density between AGN at northern (1.5 Jy) and southern (2.0 Jy) declinations to account for differences in instrument sensitivity. The high fidelity imaging of this sample makes it ideal for jet-bend measurements.  The sample is also relatively large, which is advantageous from a statistical standpoint. The data from MOJAVE were not available when Paper I was published. All MOJAVE AGN were imaged at 15 GHz in the period August 1994 to September 2007 \citep{Lister2009a}.

\section{Results \& Discussion}

\citet{vonMontigny1995} suggest two scenarios where jet bending could influence $\gamma$-ray detection in radio-loud AGN. If $\gamma$-ray emission is produced within an AGN jet, and if the jet bends significantly downstream of the $\gamma$-ray emitting region, there is a possibility that the $\gamma$-ray emission is beamed away from our line of sight (Doppler-dimmed) whilst radio emission is beamed towards us. In this case we would observe a $\gamma$-ray-quiet but radio-loud AGN. The other possibility is that there may be AGN with jets either with no bends or with bends upstream of the $\gamma$-ray-emitting region, producing aligned radio and $\gamma$-ray beaming cones. In this case, we would observe a $\gamma$-ray and radio-loud AGN. Such possibilities may account for $\gamma$-ray observations not otherwise explained by relativistic beaming.

Two other simple scenarios may also be considered. First, objects for which we lie within the $\gamma$-ray beaming cone but not the radio beaming cone. Such sources would be observable to us as $\gamma$-ray-loud, radio quiet AGN. However, it is reasonable to assume that substantial radio emission occurs in regions of $\gamma$-ray emission, making this scenario unlikely. Second, AGN jets for which both $\gamma$-ray and radio beaming cones are misaligned with our line of sight. These sources will likely not be identified in $\gamma$-rays and may appear as weak radio sources.

In our approach, the number of parsec-scale jet bends and the maximum parsec-scale jet-bending angles are recorded for AGN with VLBI images that present a discernible jet structure from which jet-bend data may be extracted. The hypothesis of \citet{vonMontigny1995} allows the possibility that jet-bending could be significant on kiloparsec scales. However, neither \citet{Bower1997} nor Paper I find a significant difference in the parsec-to-kiloparsec misalignment angle between $\gamma$-ray-loud and $\gamma$-ray-quiet populations. In addition, Paper I does not find a statistically significant difference in the number of kiloparsec-scale jet-bends. More recent investigations assert that $\gamma$-ray emission originates in the parsec-scale region, near the core, although the precise region continues to be discussed \citep{Jorstad2001,Lahteenmaki2003,Sikora2009}. In the scenario proposed in these studies, both $\gamma$-ray and radio emission may be produced far upstream of any kiloparsec-scale bending, allowing the detection of both kinds of emission regardless of the existence of kiloparsec bending. Kiloparsec-scale bending does not appear to impact $\gamma$-ray detection in AGN, consistent across a number of studies, and is not explored here.

We applied the following method to our jet-bending measurements. For a given AGN, we compared VLBI images available from the RFC and in other major surveys. Considering the resolution and common features between epochs, we determined the AGN's jet structure that could be discernable by eye. In this process, we assume that an abrupt change in the apparent direction of a jet, associated with a `shock' region of high flux density, corresponds to the bending between two jet sections. We selected the best VLBI image to derive its jet properties, with preference always given to available MOJAVE images due to their superior image quality. The initial angle of the jet from the core was measured east of north in the equatorial coordinate system. The angles of any existing bends in the jet are then measured (in the same sense) with respect to the previous jet trajectory closer to the core. The number of bends and the maximum jet-bending angle were then recorded for each source, shown in Tables \ref{tab:gl_list} and \ref{tab:gq_list}.

There are a number of factors affecting an AGN's derived jet properties. Inhomogeneity is introduced from the variable resolution and sensitivity of VLBI instrumentation, the choice of radio frequency for imaging, and different image epochs for different sets of images associated with a particular AGN. As such, the low fidelity of some VLBI images may affect the consistency of jet measurements across the whole sample. Despite this, using the best available VLBI images for each AGN will provide the most accurate jet-bending properties to retest the jet-bending hypothesis. The ability to resolve jet components also diminishes with decreasing radio flux density. In weak radio sources, this is particularly evident in the reduced ability to resolve low flux density jet components (and therefore jet bending) at large distances from the core. Deflections in jets may not be obvious at small jet-bending angles, and where there are jet-bends, they may not occur at a single, well-defined point along the jet. The comparison between multiple VLBI images of AGN across different imaging frequencies and epochs also allows us to obtain jet properties that reflect the jet structure from the observer's point of view as accurately as possible. Furthermore, bright individual features with different trajectories within or around a jet may also give the false appearance of jet bending in jet structure in certain imaging epochs \citep{Homan2012}. Comparing VLBI images at different epochs assists in identifying bright regions that may temporarily distort the appearance of the jet structure. Measurements are subject to image fidelity and morphological interpretation, however, these considerations should provide a robust result when testing the significance of jet bending in $\gamma$-ray detection compared to the findings of previous, more-limited studies.

We employ a number of statistical tests to explore differences between $\gamma$-ray-loud and $\gamma$-ray-quiet populations in our sample. A two-sided Kolmogorov-Smirnov (K-S) two-sample test was used to determine the statistical significance of differences at 4.85 GHz flux density and maximum jet-bend angle distributions between $\gamma$-ray-loud and $\gamma$-ray-quiet AGN populations. Since the values for the number of jet bends are discrete, a $\chi^{2}$ test was applied to characterize differences between the distributions in this case. 13 $\gamma$-ray-loud and 32 $\gamma$-ray-quiet sources did not show discernible jets and thus were omitted from the statistical testing of jet properties. For all tests, the difference between sample populations is considered significant for a given property when, assuming the null hypothesis represents both samples being drawn from the same parent population, the probability of observing a test statistic at least as extreme as the one obtained ($p(H_{0})$) is less than 0.05 (95\% confidence that the null hypothesis can be rejected). These results are summarized in Table \ref{tab:agn_stat}.

\subsection{The RFC-Based Sample}

\begin{figure}[!phtb]
\centering
\captionsetup[subfigure]{justification=raggedright,captionskip=0mm}
\subfloat{
 \label{fig:lfx_agn_gl}
 \includegraphics[width=80mm, height=40mm]{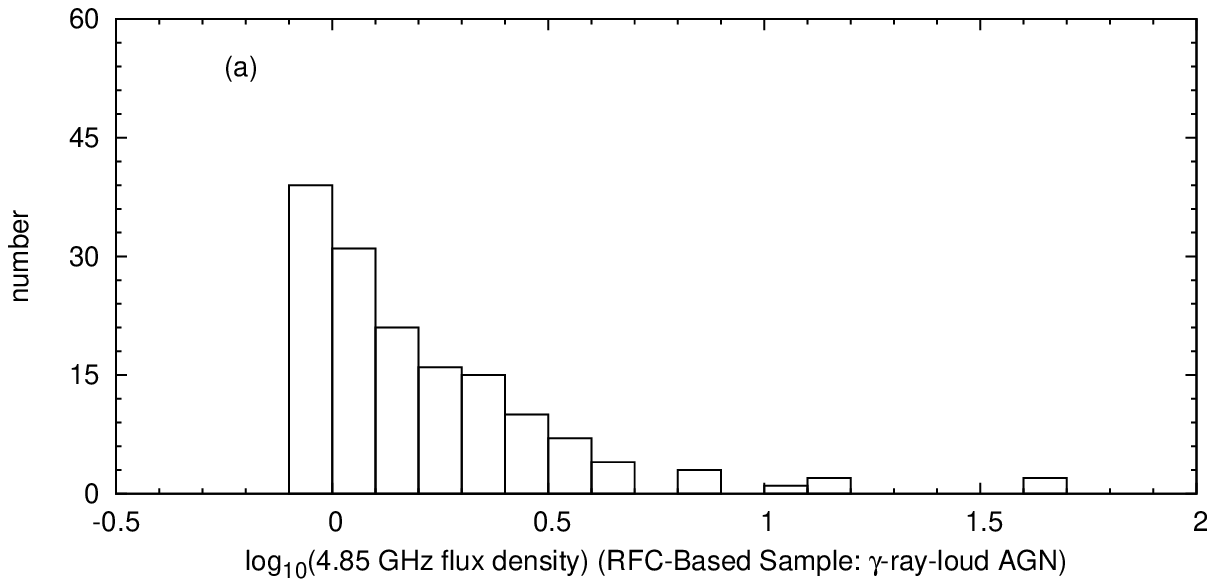}}
\subfloat{
 \label{fig:lfx_agn_gq}
 \includegraphics[width=80mm, height=40mm]{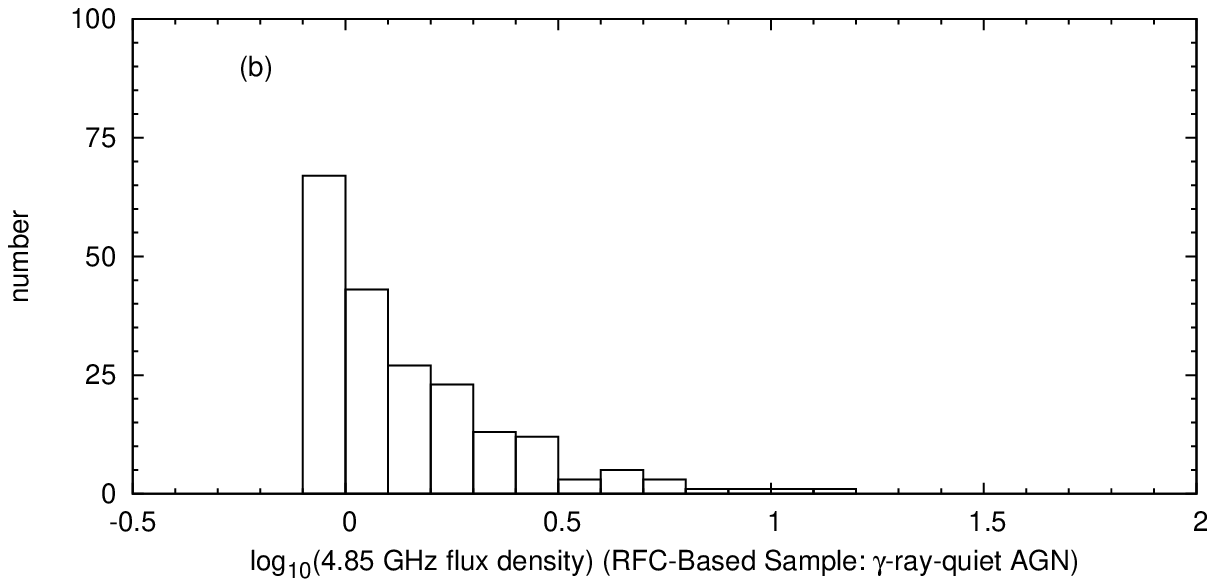}}\\
\subfloat{
 \label{fig:mb_agn_gl}
 \includegraphics[width=80mm, height=40mm]{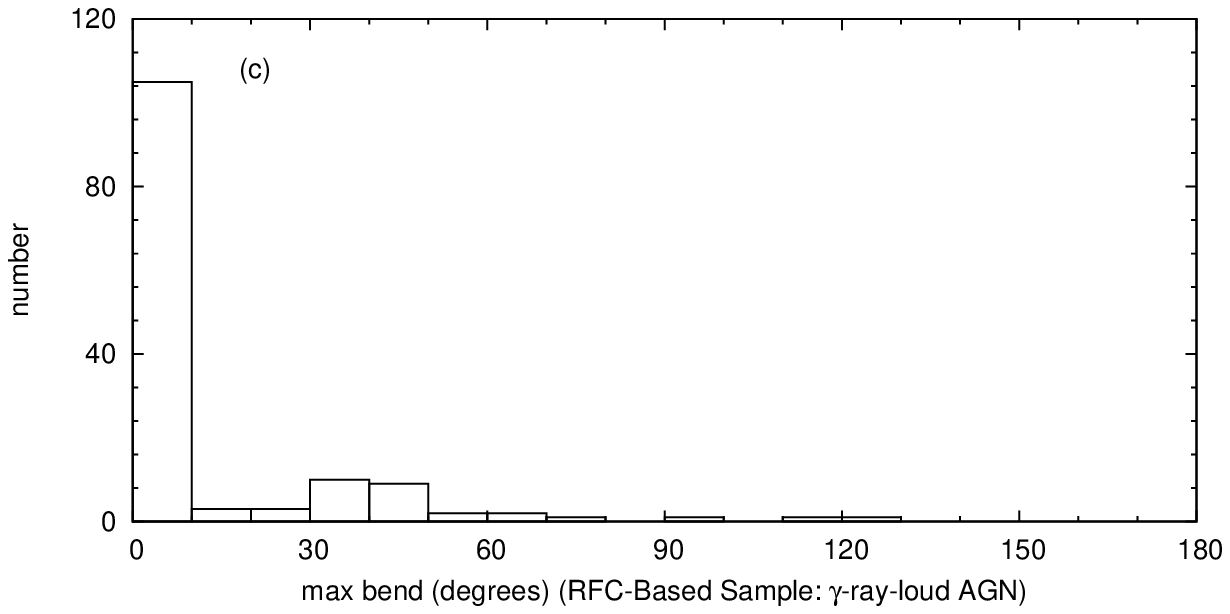}}
\subfloat{
 \label{fig:mb_agn_gq}
 \includegraphics[width=80mm, height=40mm]{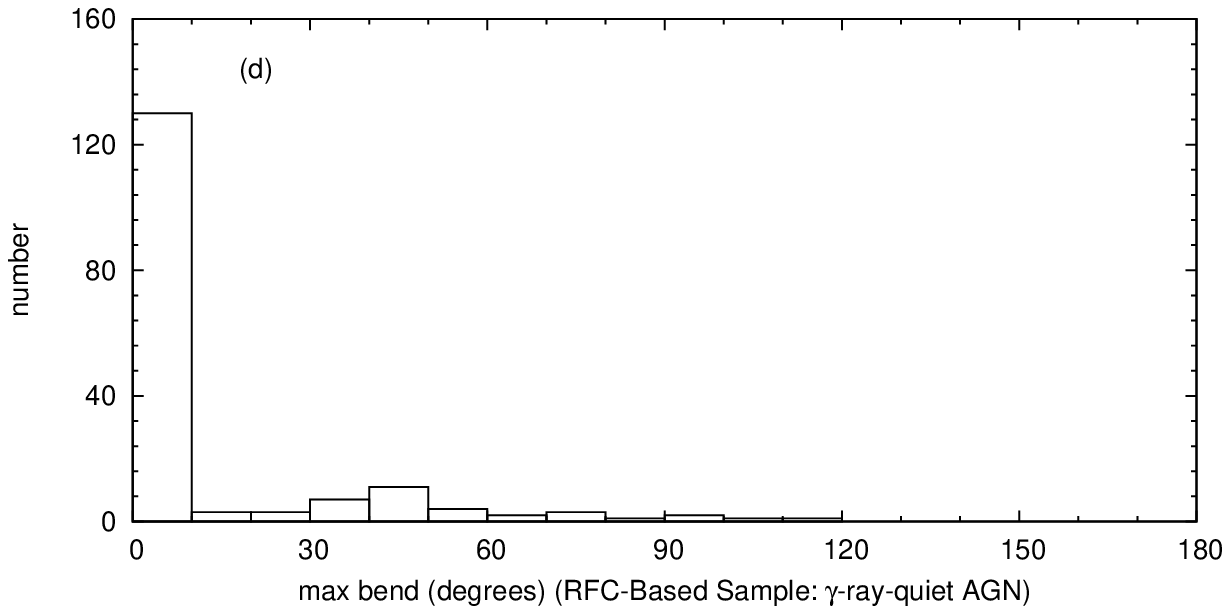}}\\
\subfloat{
 \label{fig:nb_agn_gl}
 \includegraphics[width=80mm, height=40mm]{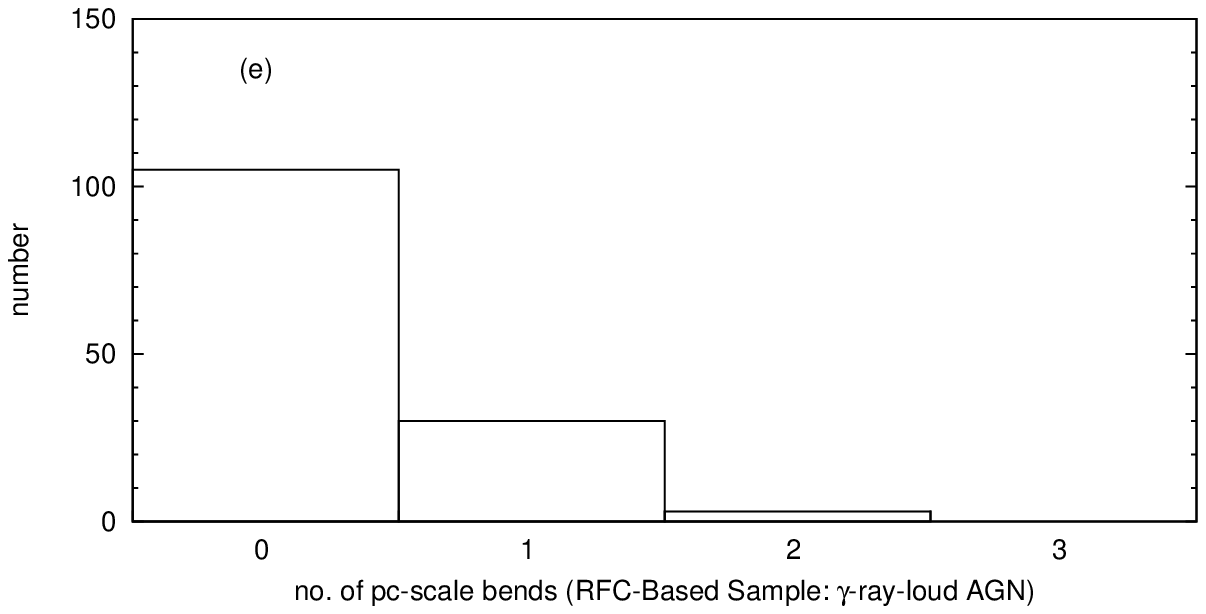}}
\subfloat{
 \label{fig:nb_agn_gq}
 \includegraphics[width=80mm, height=40mm]{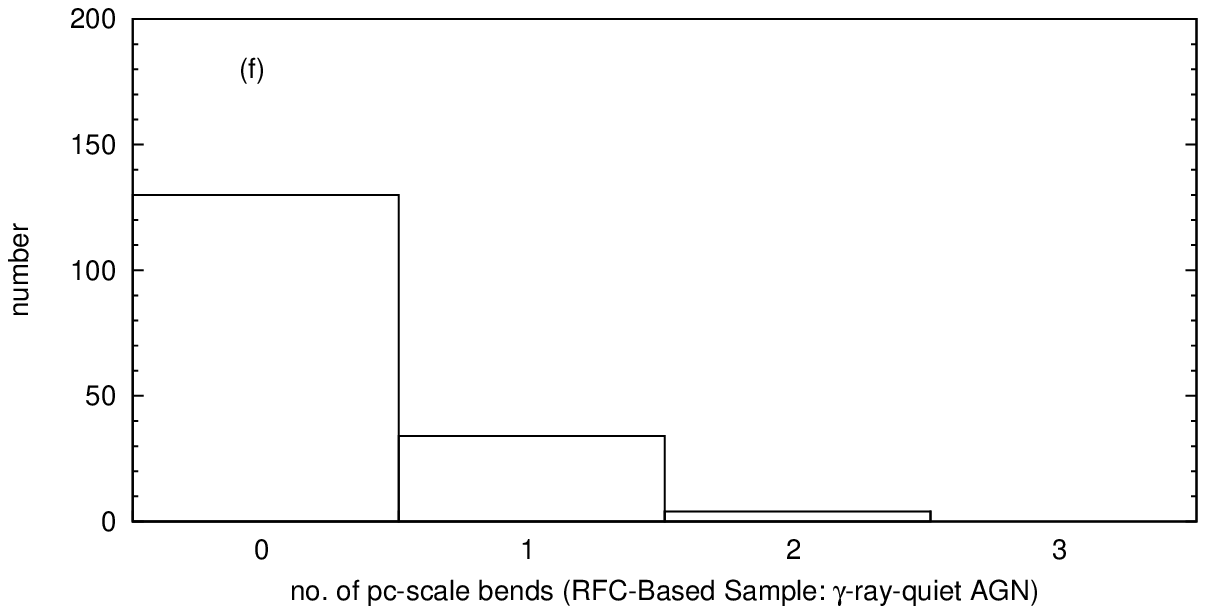}}
\caption{\label{fig:lmn_agn}Distributions of \protect\subref{fig:lfx_agn_gl}-\protect\subref{fig:lfx_agn_gq}  log$_{10}$(4.85 GHz flux density), \protect\subref{fig:mb_agn_gl}-\protect\subref{fig:mb_agn_gq} maximum bend and \protect\subref{fig:nb_agn_gl}-\protect\subref{fig:nb_agn_gq} number of bends for $\gamma$-ray loud and $\gamma$-ray-quiet radio AGN of the RFC-Based Sample.}
\end{figure}

Figures \ref{fig:lfx_agn_gl} and \ref{fig:lfx_agn_gq} show similar distributions for AGN total flux densities at 4.85 GHz. A K-S test determines that the probability that the two samples have been drawn from the same parent population is $p(H_{0}) = 0.121$, finding no statistically significant difference between the two distributions. The result gives confidence that there no biases in LAT detections that may skew jet-bending statistics.

We now consider the jet-bending statistics of $\gamma$-ray-loud and $\gamma$-ray-quiet samples. The distributions between samples for the number of jet bends (Figures \ref{fig:nb_agn_gl} and \ref{fig:nb_agn_gq}) show no significant differences. A $\chi^{2}$ test finds a statistically insignificant probability of $p(H_{0}) = 0.945$ ($\chi^{2} = 0.112$ with 2 degrees of freedom). Similarly, histograms for the parsec-scale maximum jet-bend angle (Figures \ref{fig:mb_agn_gl} and \ref{fig:mb_agn_gq}) demonstrate no significant difference between the two populations. The probability that the two samples belong to the same parent population is $p(H_{0}) = 0.998$.

\subsection{The MOJAVE Subsample}

\begin{figure}[!phtb]
\centering
\captionsetup[subfigure]{justification=raggedright,captionskip=0mm}
\subfloat{
 \label{fig:lfx_mjv_gl}
 \includegraphics[width=80mm, height=40mm]{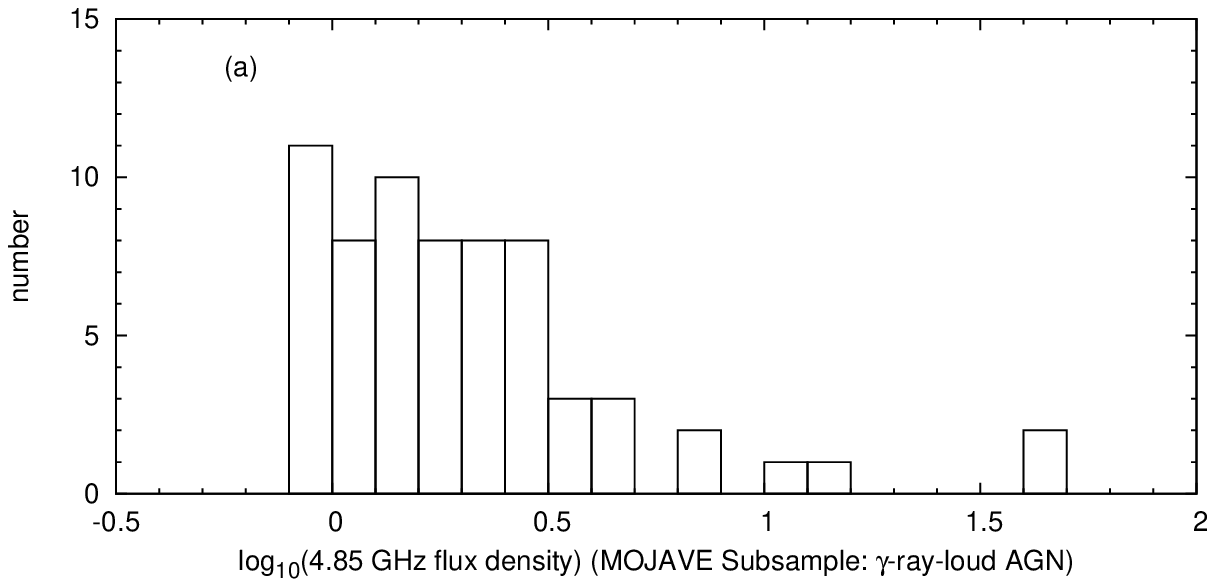}}
\subfloat{
 \label{fig:lfx_mjv_gq}
 \includegraphics[width=80mm, height=40mm]{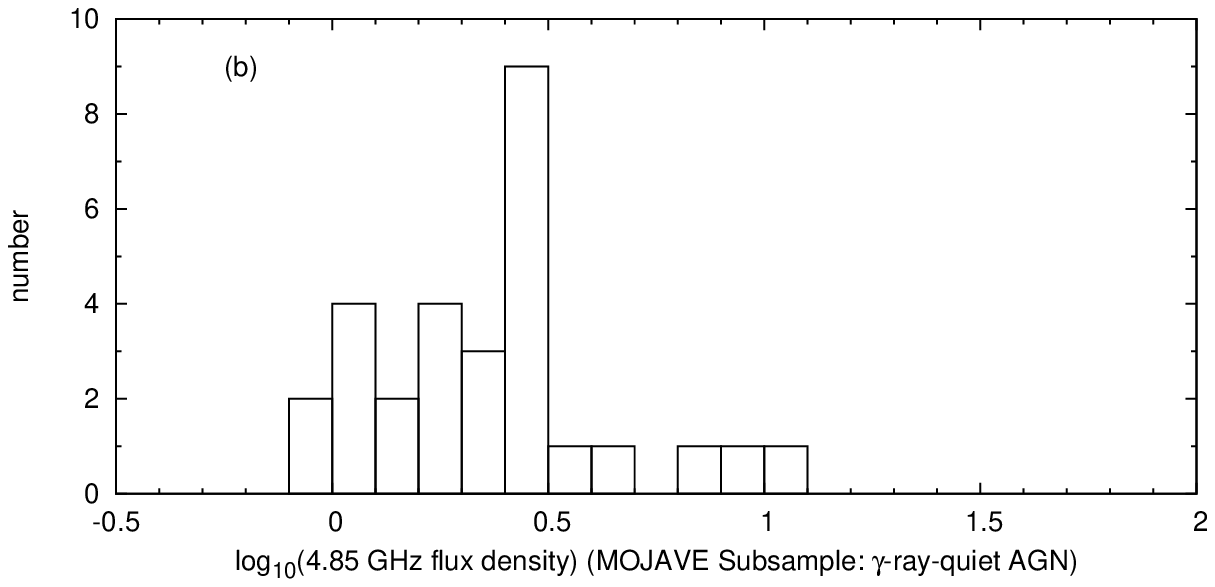}}\\
\subfloat{
 \label{fig:mb_mjv_gl}
 \includegraphics[width=80mm, height=40mm]{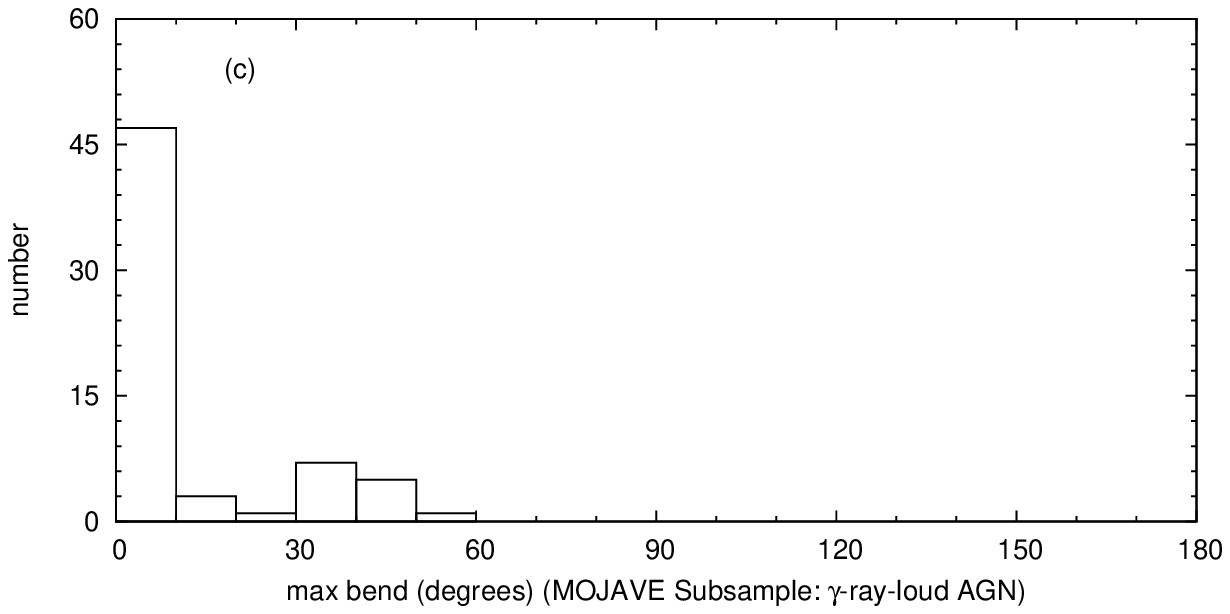}}
\subfloat{
 \label{fig:mb_mjv_gq}
 \includegraphics[width=80mm, height=40mm]{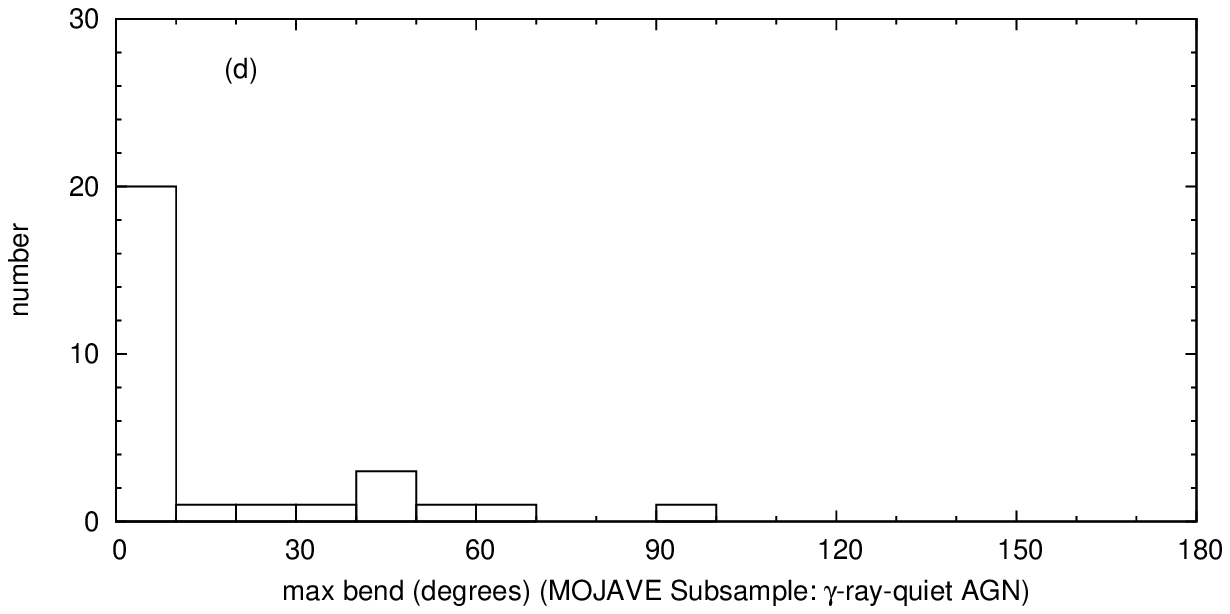}}\\
\subfloat{
 \label{fig:nb_mjv_gl}
 \includegraphics[width=80mm, height=40mm]{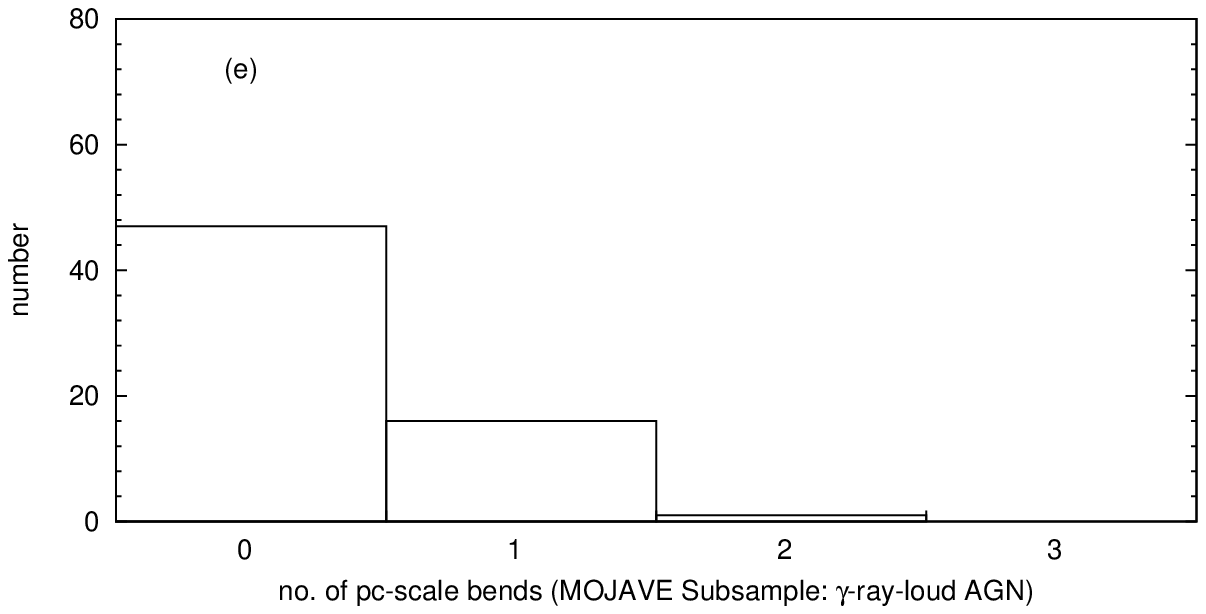}}
\subfloat{
 \label{fig:nb_mjv_gq}
 \includegraphics[width=80mm, height=40mm]{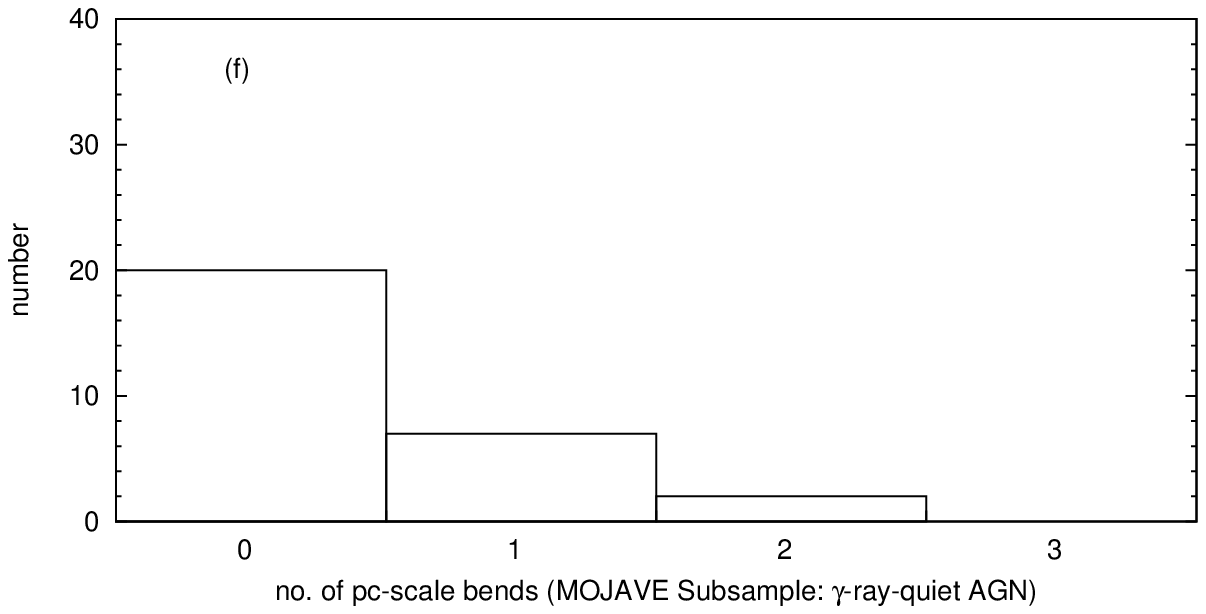}}
\caption{\label{fig:lmn_mjv}Distributions of \protect\subref{fig:lfx_mjv_gl}-\protect\subref{fig:lfx_mjv_gq}  log$_{10}$(4.85 GHz flux density), \protect\subref{fig:mb_mjv_gl}-\protect\subref{fig:mb_mjv_gq} maximum bend and \protect\subref{fig:nb_mjv_gl}-\protect\subref{fig:nb_mjv_gq} number of bends for $\gamma$-ray loud and $\gamma$-ray-quiet radio AGN of the MOJAVE Subsample}
\end{figure}

We find that there are no statistically significant differences between $\gamma$-ray-loud and $\gamma$-ray-quiet AGN populations for any of the tested properties. The results of the statistical tests on the MOJAVE Subsample are given in Table \ref{tab:agn_stat}. The $\gamma$-ray-loud and $\gamma$-ray-quiet distributions for these properties are shown in Figure \ref{fig:lmn_mjv}.

\section{Summary}

We have investigated the suggestion that $\gamma$-ray-quiet AGN have a larger tendency for jet bending than $\gamma$-ray-loud AGN \citep{vonMontigny1995}. We present a statistical analysis of a large sample of AGN that represents the population of radio-loud AGN, using data from the Fermi LAT instrument. We also conducted an analysis of a homogeneous subsample (MOJAVE) to address the inhomogeneity of selection criteria and VLBI image quality in the RFC-Based Sample.

Our analysis of the RFC-Based Sample shows that jet-bending does not play a significant role in predicting $\gamma$-ray detectability in radio-loud AGN. As such our results do not support the suggestion of \citet{vonMontigny1995}. This conclusion is in contradiction to Paper I, and supports the findings of \citet{Linford2012}. The contrast of results obtained by new data with those of Paper I reveals the strong dependence on the fidelity and homogeneity of VLBI images in jet-bend measurements. The limited quality and homogeneity of VLBI images available at the time of study appears to be the reason for the reported conclusions in Paper I.

The results of the MOJAVE Subsample generally support the conclusions drawn from the RFC-Based Sample. Statistically insignificant results were found for both the number of parsec-scale jet bends and maximum parsec-scale jet bend, indicating that the RFC-Based Sample is not adversely affected by VLBI image inhomogeneity and giving additional confidence to the null result.

\acknowledgements

We thank the referee for providing helpful comments and suggestions. This research has made use of data from the Radio Fundamental Catalog (RFC) maintained by L. Petrov, solution rfc\textunderscore2013d (unpublished, available on the Web at http://asrtogeo.org/vlbi/solutions/rfc\textunderscore2013d); the MOJAVE database, maintained by the MOJAVE team (Lister et al. 2009, AJ, 137, 3718); the CRATES catalog (Healey et al. 2007, ApJS, 171, 61); and the NASA/IPAC Extragalactic Database (NED), operated by the Jet Propulsion Laboratory, California Institute of Technology, under contract with the National Aeronautics and Space Administration.

\clearpage

\section*{Figures for RFC-Based Sample Results}

\clearpage

\section*{Figures for MOJAVE Subsample Results}

\clearpage

\begin{deluxetable}{lrcrrl}
\tabletypesize{\scriptsize}
\tablecaption{Jet Bend Data for $\gamma$-Ray-Loud Radio AGN\label{tab:gl_list}}

\tablehead{
\colhead{Source} & \colhead{$S_{4.85}$ (Jy)} & \colhead{$\phi_{\mathrm{pc}}$ ($^{\circ}$)} & \colhead{$N_{\mathrm{pc}}$} & \colhead{$|\theta_{\mathrm{pc}}|_{\mathrm{max}}$ ($^{\circ}$)} & \colhead{References}
}

\startdata
J0049$-$5738 & 1.338 & 302/$-$28 & 1 & 28 & 10, 12$^{*}$, 14 \\
J0051$-$0650 & 0.841 & 307 & 0 & 0 & 14 \\
J0116$-$1136 & 1.488 & 325 & 0 & 0 & 2, 4, 6, 14$^{*}$ \\
J0120$-$2701 & 1.000 & 156 & 0 & 0 & 6, 11, 14$^{*}$, 15 \\
J0136$+$4751 & 2.016 & 330 & 0 & 0 & 2, 4, 6, 8, 9$^{*}$, 13, 14, 18 \\
J0137$-$2430 & 0.956 & 75 & 0 & 0 & 2, 6, 11$^{*}$, 14 \\
J0141$-$0928 & 0.940 & 224 & 0 & 0 & 2, 3, 6$^{*}$, 14 \\
J0145$-$2733 & 0.833 & 53 & 0 & 0 & 14 \\
J0204$-$1701 & 1.350 & 12 & 0 & 0 & 2, 4$^{*}$, 6, 14 \\
J0204$+$1514 & 3.073 & 303 & 0 & 0 & 1, 2, 4, 6, 8, 9$^{*}$, 14, 18 \\

\enddata

\tablecomments{Columns are as follows: (1) J2000 coordinate based name; (2) total 4.85 flux density in Jy, given by the CRATES catalog \citep{Healey2007}; (3) parsec-scale jet position angles, where entries with multiple values separated by slashes denote the bending angles along the jet; (4) number of parsec-scale jet bends; (5) maximum jet-bend angle; (6) VLBI map references ($^{*}$ denotes the reference used to determine jet bends and maximum jet-bending angle where multiple references exist). Table \ref{tab:gl_list} is published in its entirety in the electronic edition of the Astrophysical Journal. A portion is shown here for guidance regarding its form and content.)}

\tablerefs{
(1) \citealp{Bondi1996}; (2) \citealp{Dodson2008} and \citealp{Scott2004}; (3) \citealp{Fey1997}; (4) \citealp{Fey2000}; (5) \citealp{Fey1996}; (6) \citealp{Fomalont2000}; (7) \citealp{Henstock1995}; (8) \citealp{Lee2008}; (9) \citealp{Lister2009a} (MOJAVE); (10) \citealp{Ojha2005}; (11) \citealp{Ojha2004}; (12) \citealp{Ojha2010}; (13) \citealp{Pearson1988}; (14) Radio Fundamental Catalog (RFC, rfc\textunderscore2013d); (15) \citealp{Shen1998}; (16) \citealp{Taylor1994}; (17) \citealp{Taylor1996}; (18) \citealp{Tingay1998} (Paper I); (19) \citealp{Tingay2002}; (20) \citealp{Xu1995}.
}

\end{deluxetable}

\clearpage

\begin{deluxetable}{lrcrrl}
\tabletypesize{\scriptsize}
\tablecaption{Jet Bend Data for $\gamma$-Ray-Quiet Radio AGN\label{tab:gq_list}}

\tablehead{
\colhead{Source} & \colhead{$S_{4.85}$ (Jy)} & \colhead{$\phi_{\mathrm{pc}}$ ($^{\circ}$)} & \colhead{$N_{\mathrm{pc}}$} & \colhead{$|\theta_{\mathrm{pc}}|_{\mathrm{max}}$ ($^{\circ}$)} & \colhead{References}
} 

\startdata

J0006$-$0623 & 2.463 & 279 & 0 & 0 & 2, 3, 6, 8, 9$^{*}$, 14 \\
J0010$+$1724 & 0.989 & 266 & 0 & 0 & 4$^{*}$, 14 \\
J0013$+$4051 & 1.035 & 331 & 0 & 0 & 2, 3, 6, 14$^{*}$, 20 \\
J0019$+$7327 & 1.583 & 144 & 0 & 0 & 2, 3, 6, 8, 9$^{*}$, 13, 14, 18 \\
J0024$-$4202 & 2.036 & 297 & 0 & 0 & 14 \\
J0029$+$3456 & 1.182 & 56 & 0 & 0 & 3, 6, 14$^{*}$, 16 \\
J0038$+$4137 & 1.144 & 113 & 0 & 0 & 6$^{*}$, 7, 14 \\
J0042$+$2320 & 1.604 & 118 & 0 & 0 & 2, 4, 6, 14$^{*}$ \\
J0051$-$4226 & 0.926 & 40 & 0 & 0 & 14 \\
J0057$+$3021 & 0.914 & 309 & 0 & 0 & 6$^{*}$, 14 \\

\enddata

\tablecomments{Columns are as follows: (1) J2000 coordinate based name; (2) total 4.85 flux density in Jy, given by the CRATES catalog \citep{Healey2007}; (3) parsec-scale jet position angles, where entries with multiple values separated by slashes denote the bending angles along the jet; (4) number of parsec-scale jet bends; (5) maximum jet-bend angle; (6) VLBI map references ($^{*}$ denotes the reference used to determine jet bends and maximum jet-bending angle where multiple references exist). Table \ref{tab:gq_list} is published in its entirety in the electronic edition of the Astrophysical Journal. A portion is shown here for guidance regarding its form and content.}

\tablerefs{
(1) \citealp{Bondi1996}; (2) \citealp{Dodson2008} and \citealp{Scott2004}; (3) \citealp{Fey1997}; (4) \citealp{Fey2000}; (5) \citealp{Fey1996}; (6) \citealp{Fomalont2000}; (7) \citealp{Henstock1995}; (8) \citealp{Lee2008}; (9) \citealp{Lister2009a}; (10) \citealp{Ojha2005}; (11) \citealp{Ojha2004}; (12) \citealp{Ojha2010}; (13) \citealp{Pearson1988}; (14) Radio Fundamental Catalog (RFC, rfc\textunderscore2013d); (15) \citealp{Shen1998}; (16) \citealp{Taylor1994}; (17) \citealp{Taylor1996}; (18) \citealp{Tingay1998} (Paper I); (19) \citealp{Tingay2002}; (20) \citealp{Xu1995}.
}

\end{deluxetable}

\clearpage

\begin{deluxetable}{p{90mm}lrl}
\tabletypesize{\footnotesize}
\tablecaption{Statistical Results (K-S and $\chi^{2}$ tests)\label{tab:agn_stat}}

\tablehead{
\colhead{Source Property} & \colhead{Statistical Test} & \colhead{$p(H_0)$} & \colhead{Significant}
}

\startdata

\sidehead{RFC-Based Sample}
Total 4.85 GHz flux density\dotfill & K-S & 0.121 & no \\
Maximum pc-scale jet bend\dotfill & K-S & 0.998 & no \\
Number of pc-scale jet bends\dotfill & $\chi^{2}$ & 0.945 & no \\

\tableline
\sidehead{\citealp{Lister2009a} (MOJAVE) Sample}
Total 4.85 GHz flux density\dotfill & K-S & 0.145 & no \\
Maximum pc-scale jet bend\dotfill & K-S & 0.948 & no \\
Number of pc-scale jet bends\dotfill & $\chi^{2}$ & 0.402 & no \\

\enddata

\tablecomments{A result is considered significant when, assuming the null hypothesis holds, the probability of observing a test statistic at least as extreme as the one obtained $p(H_{0})$ is less than 0.05 (95\% confidence level).}

\end{deluxetable}

\end{document}